\documentstyle[aasms4,12pt]{article}

\def \aa    #1 #2   {{ A\&A \/}, {#1}, {#2}}
\def \aas   #1 #2   {{ A\&AS \/}, {#1}, {#2}}
\def \aj    #1 #2   {{ AJ \/}, {#1}, {#2}}
\def \apjl  #1 #2   {{ ApJ Lett. \/}, {#1}, {#2}}
\def \apj   #1 #2   {{ ApJ \/}, {#1}, {#2}}
\def \apjs  #1 #2   {{ ApJS \/}, {#1}, {#2}}
\def \mnras #1 #2   {{ MNRAS \/}, {#1}, {#2}}
\def \prl   #1 #2   {{ Phys. Rev. Lett. \/}, {#1}, {#2}}
\def \nat   #1 #2   {{ Nature \/}, {#1}, {#2}}
\def \com   #1 #2   {{ Comments Astrophys.\/}, {#1}, {#2}}
\def \sast  #1 #2   {{ Soviet Astron. \/} {#1}, {#2}}
\def \sastl #1 #2   {{ Soviet Astron. Lett.\/}, {#1}, {#2}}
\def \astr  #1 #2   {{ Astrophysics \/}, {#1}, {#2}}

\begin{document}

\title{Detection of Network Structure in the Las Campanas Redshift Survey}

\author{Sergei F. Shandarin $\&$ Capp Yess\\
Department of Physics and Astronomy \\
University of Kansas, Lawrence, KS 66045}

\begin{abstract}

We employ a percolation technique developed for pointwise distributions to 
analyze two-dimensional projections of the three northern and three southern 
slices in the Las Campanas Redshift Survey.  
One of the  goals of this paper is to compare 
the visual impressions of the structure within distributions with 
objective statistical analysis. 
We track the 
growth of the largest cluster as an indicator of the network structure.
We restrict our analysis to volume limited subsamples in the regions
from 200 to 400 $h^{-1}$ Mpc where the number density of
galaxies is the highest. 
As a major result, we report a measurement of an unambiguous signal, 
with high signal-to-noise ratio (at least at the level of a few $\sigma$), 
indicating significant connectivity of the galaxy distribution
which in two dimensions is indicative of a 
filamentary distribution. This is in general agreement with the 
visual impression and typical for the standard theory of the 
large-scale structure formation based on gravitational instability of
initially Gaussian density fluctuations.  

{\it Subject headings:} large-scale structure of the universe:
 observations - methods: statistical

\end{abstract}

\section{Introduction}

For decades cosmologists have been developing methods for characterizing and 
quantifying the geometry and topology of structure in the local 
galaxy distribution as supplied to 
them by astronomers. Numerous statistics have been employed and refined 
in this endeavor with the most successful being percolation analysis 
(\cite{Sh-Z83}; \cite{Ein-Kly-Saa-Sh84}) and the genus 
statistic 
(\cite{Got-Mel-Dic86}, \cite{Pea-etal97})
\footnote{As reported in a recent paper by
Bhavsar \& Splinter (1996) the percolation properties can be reconstructed from
the minimal spanning tree.}.
A more general approach that in principle accommodates both percolation and
genus statistics is based on measuring the Minkowski functionals 
(\cite{Mec-Buc-Wag94}; \cite{Sch-Buc97}). The filling factor and the genus 
(which apply to the entire distribution) are 
two of the four Minkowski functionals ($M_0, M_3$ correspondingly), 
and the volume of the largest cluster 
statistic is another ($M_0$ for
the largest cluster only).

By differing techniques, these statistics have produced compatible results 
describing the structure of the local universe in the IRAS 1.2 Jy Survey 
(\cite{Yes-Sh-Fis97}; \cite{Pro-Wei97}). Percolation analysis gives similar
results for Gaussian distributions but significantly differs from 
the genus statistic for the nonGaussian distributions (\cite{Sah-Sat-Sh97}).  
In studies of geometry and topology, 
the major limiting factors were the shot noise in the analysis of pointwise 
distributions, or resolution in the analysis of density fields derived from 
galaxy positions (\cite{Yes-Sh-Fis97}) and the small size of the survey.  
The relatively high galaxy number density  in 
the Las Campanas Redshift Survey (LCRS) 
reduces the discreteness effects.  Also, the 
size of  the survey promises that a fair sample of the 
universe is being probed. At least visually there are no structures comparable 
to the size of the slices. For the first time a redshift survey
has reached the scale where the universe looks roughly homogeneous
apart from the obvious inhomogeneity of a magnitude limited survey.
The extent of the upcoming Sloan Digital Sky Survey 
(see e.g. \cite{Gun-Wei95}) promises equally unequivocal results over even 
larger regions.  

The particulars of the Las Campanas Redshift Survey and our utilization of 
the survey are detailed in $\S$ 2 of this paper.  In addition, the standard 
Poisson distributions and their application are explained in $\S$ 2.  The 
parameters used to characterize the galaxy distributions are described 
in $\S$ 3 along with the percolation method for pointwise distributions.  In 
$\S$ 4 the percolation results are presented and explained.  Conclusions are 
also drawn in $\S$ 4 with suggestions for further investigations.

\section{The LCRS and Poisson Standards}

There are approximately 25,000 galaxies with redshift positions in the LCRS.  
They are distributed over six slices, three northern and three southern.  The 
geometry of slices, schematically depicted in Figure 1, are strips of the sky 
$1.5^{\circ}$ thick and 80$^{\circ}$ wide which are 
separated by 3$^{\circ}$.  The northern slices are centered at declinations of 
-3$^{\circ}$ (N1), -6$^{\circ}$ (N2), and -12$^{\circ}$ (N3) and the southern 
slices at -39$^{\circ}$ (S1), -42$^{\circ}$ (S2), and -45$^{\circ}$ (S3).  All 
slices are probed to a depth of 60,000 km s$^{-1}$ (600 h$^{-1}$ Mpc for 
H$_{\circ}$= h km s$^{-1}$ Mpc$^{-1}$) for 
galaxies of m= 17.75, the limiting magnitude.  For the details of the LCRS and 
clustering properties see Shectman et al. (1996) and references therein.  
The survey volume of the LCRS allows for a fair assessment of the connectivity 
of the galaxy distribution given that cosmic 
structures are on the scale of 100 h$^{-1}$ Mpc and the number of galaxies 
contained in the survey gives a signal that, given the sensitivity of 
percolation analysis, overcomes random noise for a large portion of the 
survey volume.  

In order to characterize the topologies of the Las Campanas' slices, 
standards typifying random distributions need to be constructed.  
Since the galaxy positions of the slices are projected to a central conical
surface of each slice, the 
standards have to account for the local galaxy number density of the LCRS  
as well as the projection effects.  

Figure 1 schematically shows the central conical surfaces of the six slices of
the LCRS. The northern slices are nearly flat and the southern ones are 
significantly curved. We analyze the distributions obtained by projecting the
galaxies on these surfaces. Since the conical surface does not have inner 
curvature it can be unrolled onto a plane surface without distortion 
(shown at the top of Figure 1).

Figure 2 shows the two-dimensional flat map 
of a northern slice, N1, of the LCRS in the upper left panel.  
Its corresponding Poisson distributions, with corrections sequentially 
applied, are also shown to illustrate the effect of each correction: 
the projection increases the 2D density toward the outer 
border (the  bottom right panel) and the selection effect toward 
the central part of the slice (the  bottom left panel).  
Poisson distributions created to correspond to an 
appropriate selection function and corrected for projection effects serve 
as standards for pointwise distributions.  
It is worth noting that all panels in Figure 2 have the same number of points.
The selection function chosen 
to approximate the distributions of the six slices was taken from Lin  
et al. (1996) and based on the subsample of galaxies from both north and 
south slices termed NS112.  Note it is easy for the eye to discern the survey 
distribution from the random distributions.

\section{Percolation}

The first step in the percolation of pointwise distributions is to 
superimpose a grid on the sector geometry and then locate the galaxy positions 
on the lattice.  In projecting the galaxy positions to a flattened, central 
plane, we have accentually ignored the effects of curvature associated with 
the geometry of the initial survey which may possibly have a greater
consequence for the southern slices since they are more strongly curved.  
Our percolation analysis is performed 
on a two dimensional lattice of cells 1 
Mpc${^2}$ in area and the same size for all slices and standards.  The 
positions of galaxies are equated with filled lattice cells.  Filled cells 
that share a common side are considered neighbors.  Through the stipulation 
that `any neighbor of my neighbor is my neighbor' clusters \footnote{The term 
cluster as used in percolation analysis does not imply cluster of 
galaxies in the astronomical sense.} composed of 
adjacent cells are defined and grow.  Our percolation method for pointwise 
distributions allows for two means by which clusters can grow.  Circles of 
specified radius are constructed around the initially filled cells 
(galaxy positions) in the distribution.  The radii 
of these circles are incrementally increased to encompass adjacent cites.  
Cells enveloped by expanding circles are labeled filled and are considered 
neighbors of the initial cell at the center of the circle.  If two or more 
circles come to overlap while expanding, the members of the overlapping 
clusters merge into a single combined cluster.  As the radii increase clusters 
will grow in size and generally diminish in number due to mergers.  This 
process will continue until the largest cluster is the only cluster in the
distribution. In an infinite space, the largest cluster emerges as the 
infinite cluster. For details of the pointwise percolation method see 
Klypin $\&$ Shandarin (1993). 

In this study, we track two percolation parameters as functions of the 
increasing circle radius. The filling factor (FF) is defined as the 
fraction of filled cells in the total area.  The second parameter, the Largest
Cluster Statistic (LCS), is the 
relative size of the largest cluster to the total area of filled cells.  
Because the size of the largest cluster is reported in units of 
the FF, its initial value should be small when the 
largest cluster is one of many clusters, and it's 
maximum value is 1.0 when the largest cluster spans the space and incorporates 
all the filled cells of the distribution.  The relative area of the largest 
cluster is reported as a function of the FF in comparisons 
between galaxy and Poisson distributions (see Figure 3).

A rapid rise in the Largest Cluster Statistic is indicative of the 
percolation 
condition \footnote{The maximum rate of the growth of the largest cluster
volume can be used as an indicator of the percolation threshold, see e.g.
\cite{deL-Gel-Huc91}}(\cite{Kly-Sh93}).  
However, it is not important for this study to 
determine the exact ff$_{p}$ associated with percolation. The exact 
ff$_{p}$ is a noisier statistic than the LCS  and 
less discriminating. In our method, the LCS, 
over its full range, is used to characterize the nature of the 
distribution (\cite{Yes-Sh96}).  In general, the faster the LCS
grows, the more the curve shifts to the left and the more connected 
the distribution.  A distribution for which the 
LCS grows more rapidly than a comparable Poisson model 
is described as an example of a network topology, and a distribution for which 
the LCS grows more slowly is considered to be clumpy or 
have a `meatball' topology.  

For a direct comparison of percolation results, pointwise distribution
standards need to have equivalent
initial filling factors as the distributions they are characterizing.  
If the initial FF of a standard distribution
is to high (ff$_{\circ}$ $\gtrsim$ 0.1), the resolution of the percolation
parameters will not be sufficient to detect the onset of percolation.  
The Poisson standards
in this study are random distributions adjusted, as a 
function of radius, for selection and projection effects with corresponding 
initial filling factors well below the resolution limit.
The intial filling factors of the survey slices fall in the range 1-3\% which 
are much smaller than the percolation transitions (see Figure 3).
In addition, it is easy to see that because of our percolation method 
the resolution of the LCS can depend strongly on the value 
of the initial FF.  For instance, for distributions that initially 
have well isolated galaxies the majority of the clusters will contain only 
one filled cell.  In this case, the first iteration of the expanding circles 
will add four nearest neighbors to virtually every cluster causing the FF 
to increase by a factor of nearly five.  The initial filling factors of 
the Las Campanas slices are well below the level where this effect would 
negate the results.

\section{Results and Conclusions}

Shown in Figure 3 are the results of the Las Campanas percolation
analysis.  The left column shows the results for the three northern
slices of the survey while the right column shows the results of the
southern slices. We present the results separately as the North and 
South slices have significantly different geometries (see Figure 1) and
also different patterns of regions observed with 50-fiber
and 112-fiber spectrographs.  The southern slices have a quasiperiodic
pattern of differently observed fields.

The solid lines are the results for the survey slices with 
the lightest being N1 (S1) and the heaviest N3 (S3).
In all graphs, the dotted line to the far right is the
result from a statistically {\it homogeneous} 2D Poisson distribution with 
the appropriate initial
FF and geometry (see Figure 2, the top right panel).  
This result is shown for reference.
The dashed line is the result from an appropriate 3D Poisson
distribution projected on the 2D conical surface and
corrected for selection effects (see Figure 2, the bottom left panel). In
all cases the error bars on the reference curves represent one sigma 
deviations over four realizations.

In the top panels of Figure 3 the corrected Poisson results are 
indistinguishable from
the survey results.  An explanation for this is that the distortion caused by 
the selection function produces  statistically  inhomogeneous 
distribution. 
Mixing the regions with very different number densities of galaxies
at the first glance looks like the shift to the left but it completely
smears out the characteristic percolation transition. 
As a result the LCS curve looks like a featureless almost straight line.
Thus, we conclude  the percolation
analysis requires more statistically homogeneous sampling than geometrically 
blind statistics.
 
By separating the survey into two regions ( 60 $\leq R \leq$ 400 h$^{-1}$ Mpc
and 400 $\leq R \leq$ 600 h$^{-1}$ Mpc ) the effect of the selection function
is considerably reduced and the LCS of the Poisson distribution 
shifts to the right and acquires a characteristic form (see Figure 3, the 
middle panels).  The
reason is that in a magnitude limited survey the selection function is
dependent on the distance from the observer and in the case of the LCRS peaks
at a value of $R \approx$ 200 h$^{-1}$ Mpc.  
The middle panels show the results of a magnitude limited sample in the region 
60 $\leq R \leq$ 400 h$^{-1}$ Mpc.  There is now a clear distinction 
between survey and corrected Poisson distribution results especially for the
North slices.  Results for the
region 400 $\leq R \leq$ 600 h$^{-1}$ Mpc (not shown) are qualitatively 
similar to those for the inner region shown.  The slices 
percolate at lower filling factors than the random standard in all cases.
More homogeneous subsamples of the survey reveal a clear and
unambiguous signal for a connected topology in the region analyzed.

The bottom panels show the results for volume limited subsamples 
(see Figure 4)
derived from the survey in the region 200 $\leq R \leq$ 400 h$^{-1}$ Mpc. 
\footnote{ Volume limited samples with the same geometry as the 
magnitude limited samples were also analyzed.  The results for those samples
where similar to the results shown and the conclusions are the same.}
We have restricted our analysis to the most dense central regions of
the slices where the combined selection and projection effects are the least.
Once again there is a clear 
signal at a few sigma level for a connected topology for both the North and
South subsamples.  
The results above illustrate the necessity for volume limited samples.
Our method contrasts with the method of de Lapparent, 
Geller and Huchra (1991) to remove the distortion effects of the selections 
function in their work with the CfA slices.  They increase the size of outer 
lying grid cells in order to
maintain a similar mean number of galaxies per grid cell throughout 
the survey.  In essence, they may
have traded one distortion for another; certainly the effects of their 
method are not understood and a comparison with their results is difficult.
They also employed a nonstandard definition of a neighbor.
  
Sources of distortion in this study are statistically inhomogeneous 
distributions due to the selection function, two-dimensional analysis of 
three-dimensional surveys, the curvature of the slices and possible inherent 
galaxy incompleteness in the survey design.  We can substantially reduce
the 
selection effects by analyzing volume limited surveys and the projection
effects by generating random two-dimensional reference catalogs that are 
similar to the surveys.  Even then, the volume limited northern slices
show somewhat stronger connectivity than the southern slices when the LCS 
is examined over its entire range due either to curvature
effects or 
regional variation in the topology at these scales.  A three dimensional study
may determine the cause of this difference.  

The results of
this study imply a connected topology for all slices consistent
with a filamentary geometry.  
However, in a survey having the thin slice geometry distinguishing between
filaments in 3D and pancakes may be intrinsically difficult.
A recent two-dimensional analysis of the 
LCRS found that the smoothed density distribution has a genus consistent with 
a random-phase Gaussian distribution of initial density fluctuations 
(\cite{Col97}).  
It has previously been shown that (\cite{Dom-Sh92}) the percolation results 
from highly smoothed, nonlinear, density distributions obtained from N-body 
simulations are consistent with Gaussian fields.
Thus, the results of the percolation 
analysis of the Las Campanas galaxy survey 
are in agreement with the standard model of structure formation 
(based on the gravitational instability and initial 
Gaussian perturbations of the density field) that arises in the inflationary 
scenarios.  

In Figure 4, the volume limited distributions for 
each slice are pictured.  It is interesting to note that the LCS 
is not predetermined by the number density  of galaxies in the
volume limited subsamples of the LCRS. 
The middle northern slice, N2, is obviously the sparsest of the 
six slices and yet it has a similar degree of connectivity as 
the other slices.  
Our experience is that visual inspection is a sensitive method for detecting
the network structures. However, it is not appropriate 
method by which to measure
the degree of connectivity or type of structures in a distribution.  
In particular, the visual impression strongly depends on the size of the
dots used in plots. In some sense percolation analysis corresponds to
scanning through the entire range of dot sizes.
There is 
also the important result that the northern slices demonstrate stronger 
connectivity than the southern slices through percolation though this is 
hard to detect by visual inspection, however the difference in LCS is
marginal and may be purely statistical.  Another important aspect of the Las 
Campanas survey is that there are small radial gaps in some of the slices.  
These radial voids are a cause a systematic error in the analysis and 
act to shift the LCS to the right.  The percolation 
analysis under these 
circumstances gives a lower estimate of the degree of connectivity.  

We intend to expand 
our study to three dimensions with a complete analysis of uncertainties.  
In order to quantify the degree of connectivity in the survey, an assessment of 
edge and grid orientation effects must be made.  Landy et al. (1996) 
have reported an
enhancement of the power spectrum on length scales of roughly 100 h$^{-1}$ Mpc.
This signal is associated with identifiable structures in the survey and is
highly directional.  Any effects on percolation results due to the 
directionality of the k-space fluctuations needs to be assessed by rotating 
the grid.   Also, the consequences of 
possible redshift distortions must be evaluated and incorporated into the 
results.  Fingers of God distortions are well known to astronomers but recent
work by Praton and Melott (1997) has examined distortions that result in linear
structures perpendicular to the line of sight.  A comparison between
percolation of simulated galaxy catalogs in real space and their redshift space
counterparts may help address this problem.  In addition, the resolution of 
the LCRS must be optimized in terms of grid cell size and uncertainties in 
the redshift positions.  We intend to report on these studies in a 
forthcoming paper.

The major result of this paper is the detection of the significant 
signal indicating the network structures in the galaxy distribution.
We do not compare it with the predictions of cosmological scenarios.
Those who are interested in such a comparison must generate the model
catalogs similar to the the LCRS, and we can analyze them or provide
the necessary software.

We acknowledge the support of NASA grant NAG 5-4039 and EPSCoR 1996 grant.


\newpage
\begin{thebibliography}{}


\bibitem[Bhavsar \& Splinter 1996]{Bha-Spl96}
Bhavsar, S. P., \& Splinter, R. J. 1996 \mnras {282} {1461}
\bibitem[Colley 1997]{Col97}
Colley, W. N. 1997, astro-ph/9612106
\bibitem[de Lapparent, Geller \& Huchra 1991]{deL-Gel-Huc91}
de Lapparent, V., Geller, M. J., \& Huchra, J. P. 1991 \apj {369} {273}
\bibitem[Dominirk \& Shandarin 1992]{Dom-Sh92}
Dominik, K., \& Shandarin, S.F. 1992, \apj {393} {450}
\bibitem[Einasto et al. 1984]{Ein-Kly-Saa-Sh84}
Einasto, J., Klypin, A. A., Saar, E., \& Shandarin, S. F. 1984,
\mnras {206} {529}
\bibitem[Gott, Melott \& Dickinson 1986]{Got-Mel-Dic86}
Gott, J. R., Melott, A. L., \& Dickinson, M. 1986, \apj {306} {341}
\bibitem[Gunn \& Weinberg 1995]{Gun-Wei95}
Gunn, J. E. \& Weinberg, D. H. 1995, {\it Wide-Field Spectroscopy and the
Distant Universe, eds}. S. J. Maddox and A. Arag$\acute{o}$n-Salamanca,
(Singapore : World Scientific).
\bibitem[Klypin \& Shandarin 1993]{Kly-Sh93}
Klypin, A. A., \& Shandarin, S. F. 1993, \apj {413} {48}
\bibitem[Landy et al. 1996]{Lan-etal96}
Landy, S. D., Shectman, S. A., Lin, H., Kirshner, R. P., Oemler, A. A.,
\& Tucker, D. 1996 \apjl {456} {1}
\bibitem[Lin et al. 1996]{Lin-etal96}
Lin, H., Kirshner, R. P., Shectman, S. A., Landy, S. D., Oemler, A., Tucher, D. L., \& Schechter, P. L. 1996a, ApJ, in press
\bibitem[Mecke, Buchert \& Wagner 1994]{Mec-Buc-Wag94}
Mecke, K., Buchert, T., \& Wagner, H. 1994, AA, 288, 697
\bibitem[Pearson et al 1997]{Pea-etal97}
Pearson, R.C., Coles, P., Borgani, S., Plionis, M., \& Moscardini, L.
1997, preprint
\bibitem[Praton, Melott \& McKee 1997]{Pra-Mel-McK97}
Praton, E. A., Melott, A. L., \& McKee, M. Q. 1997, 
\bibitem[Protogeros \& Weinberg 1997]{Pro-Wei97}
Protogeros, Z. A. M., \& Weinberg, D. H. 1997, astro-ph/9701147
\bibitem[Sahni, Sathyaprakash \& Shandarin 1997]{Sah-Sat-Sh97}
Sahni, V., Sathyaprakash, B. S., \& Shandarin, S. F. 1997,\apj,476, L1
\bibitem[Schmalzing \& Buchert 1997]{Sch-Buc97}
Schmalzing, J., \& Buchert, T. 1997, astro-ph/9702130
\bibitem[Shandarin \& Zel'dovich 1983]{Sh-Z83}
Shandarin, S. F., \& Zel'dovich, Ya. B. 1983, \com {10} {33}
\bibitem[Shectman et al. 1996]{She-etal96}
Shectman, S. A., Landy, S. D., Oemler, A., Tucker, D. L., Kirshner,
R. P., Lin, H., \& Schechter, P. L. 1996, {\it Wide-Field
Spectroscopy and the Distant Universe, proceedings of the 35th
Herstmonceux Conference}, eds. S. J. Maddox \& A Arag$\acute{o}$n-Salamanca,
World Scientific, Singapore
\bibitem[Yess \& Shandarin 1996]{Yes-Sh96}
Yess, C., \& Shandarin S. F. 1996, \apj {465} {2}
\bibitem[Yess, Shandarin \& Fisher 1997]{Yes-Sh-Fis97}
Yess, C., Shandarin, S. F., \& Fisher K. B. 1997, \apj {474} {553}


\end{thebibliography}
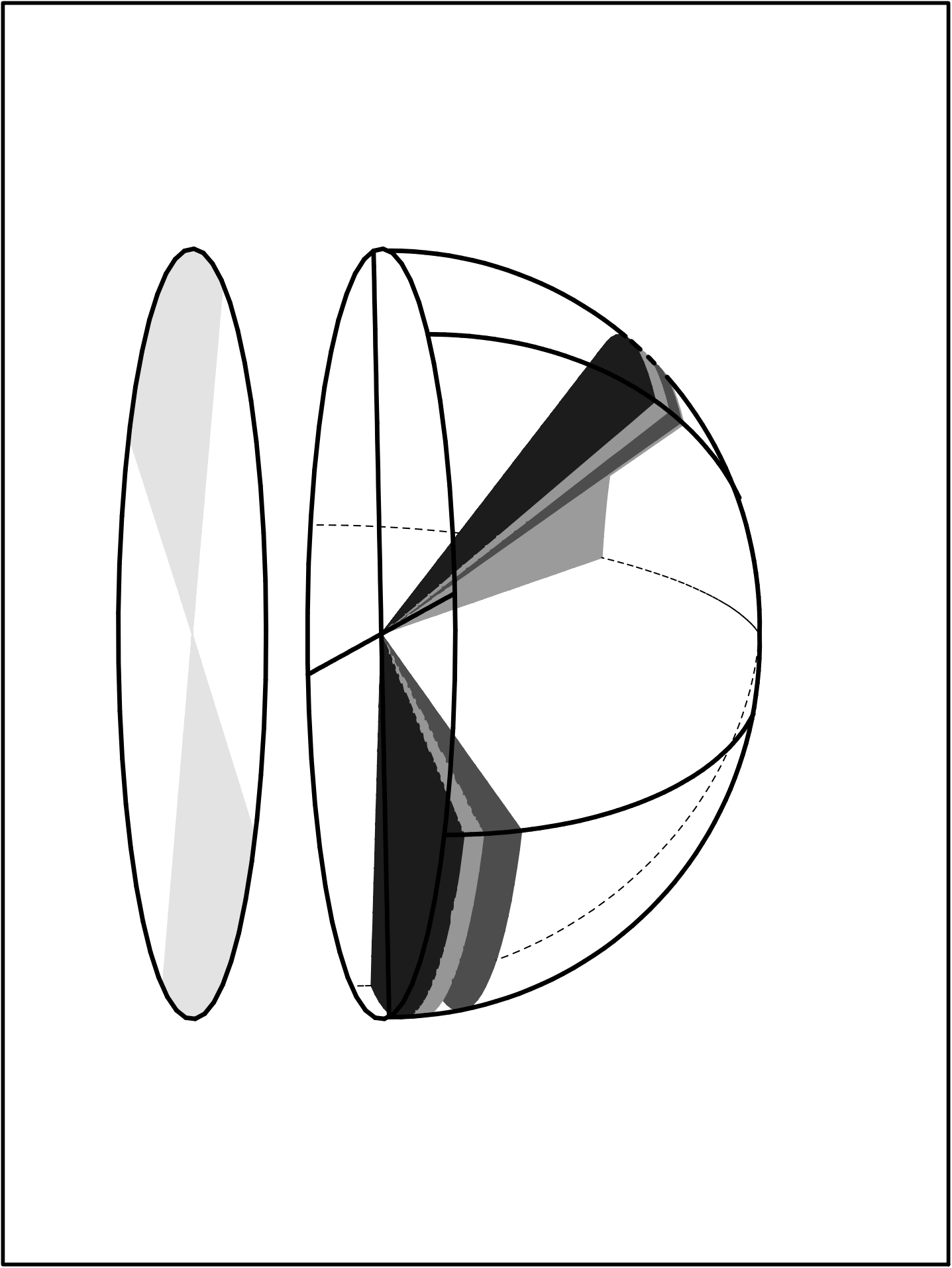
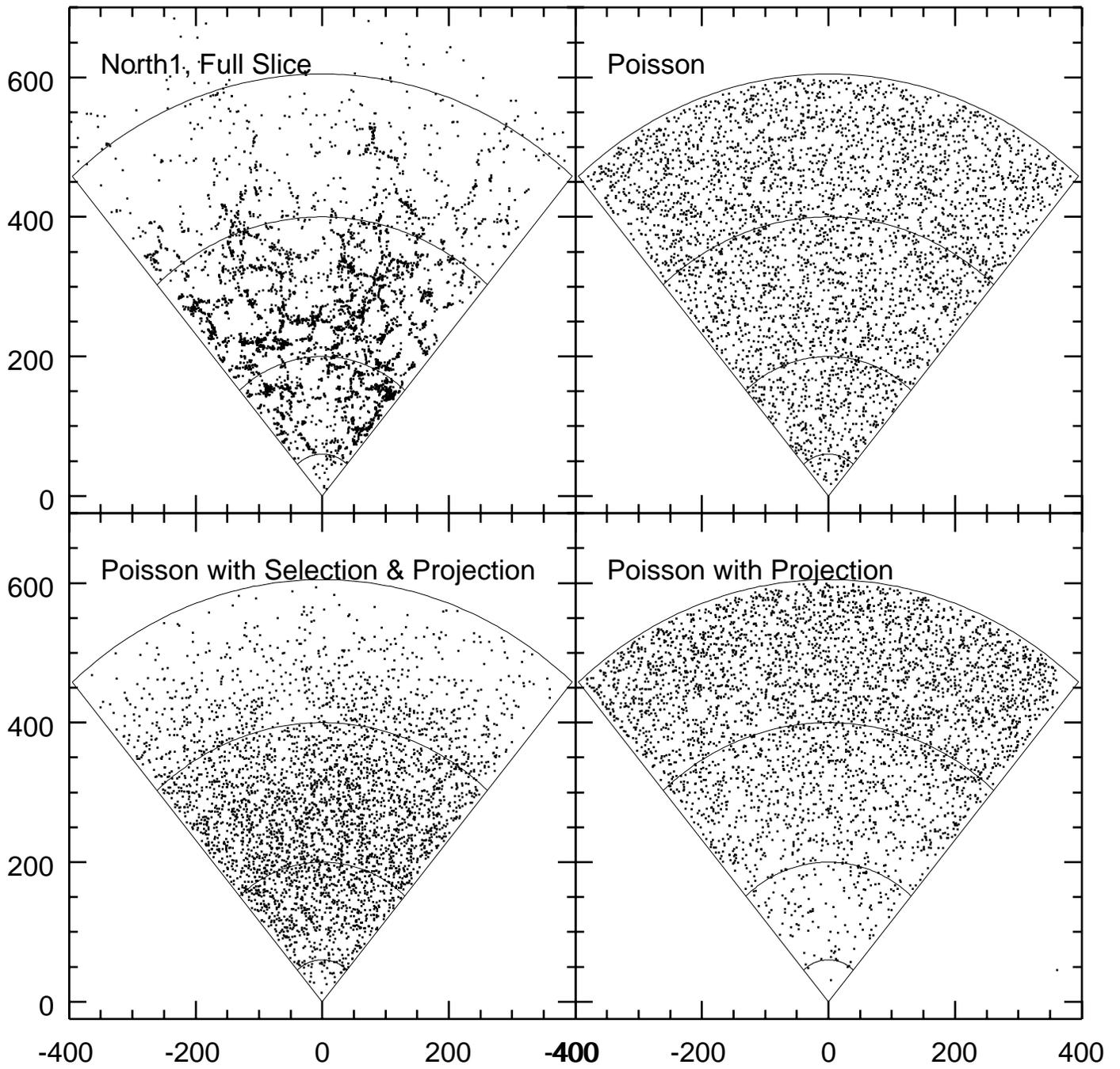
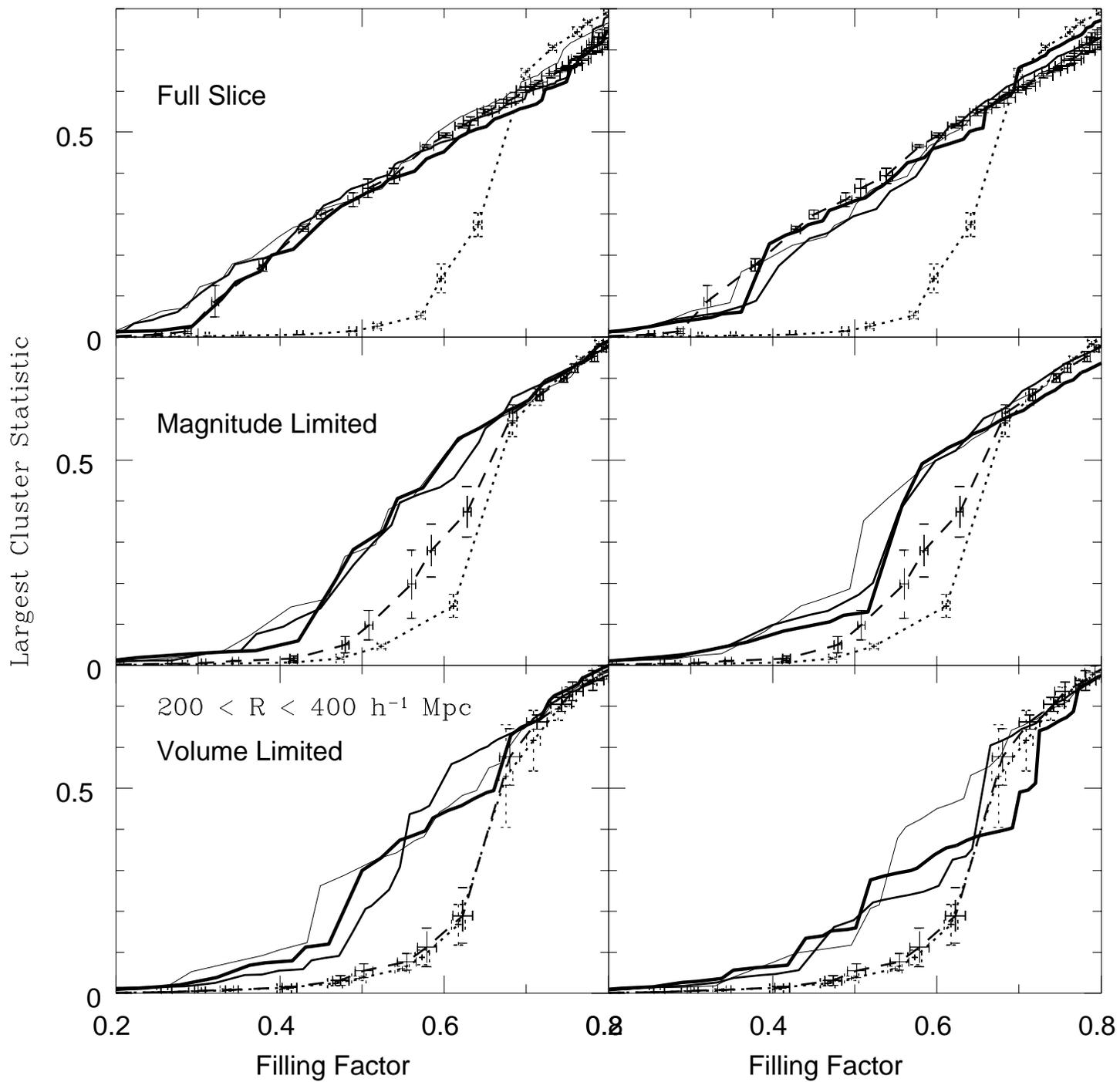
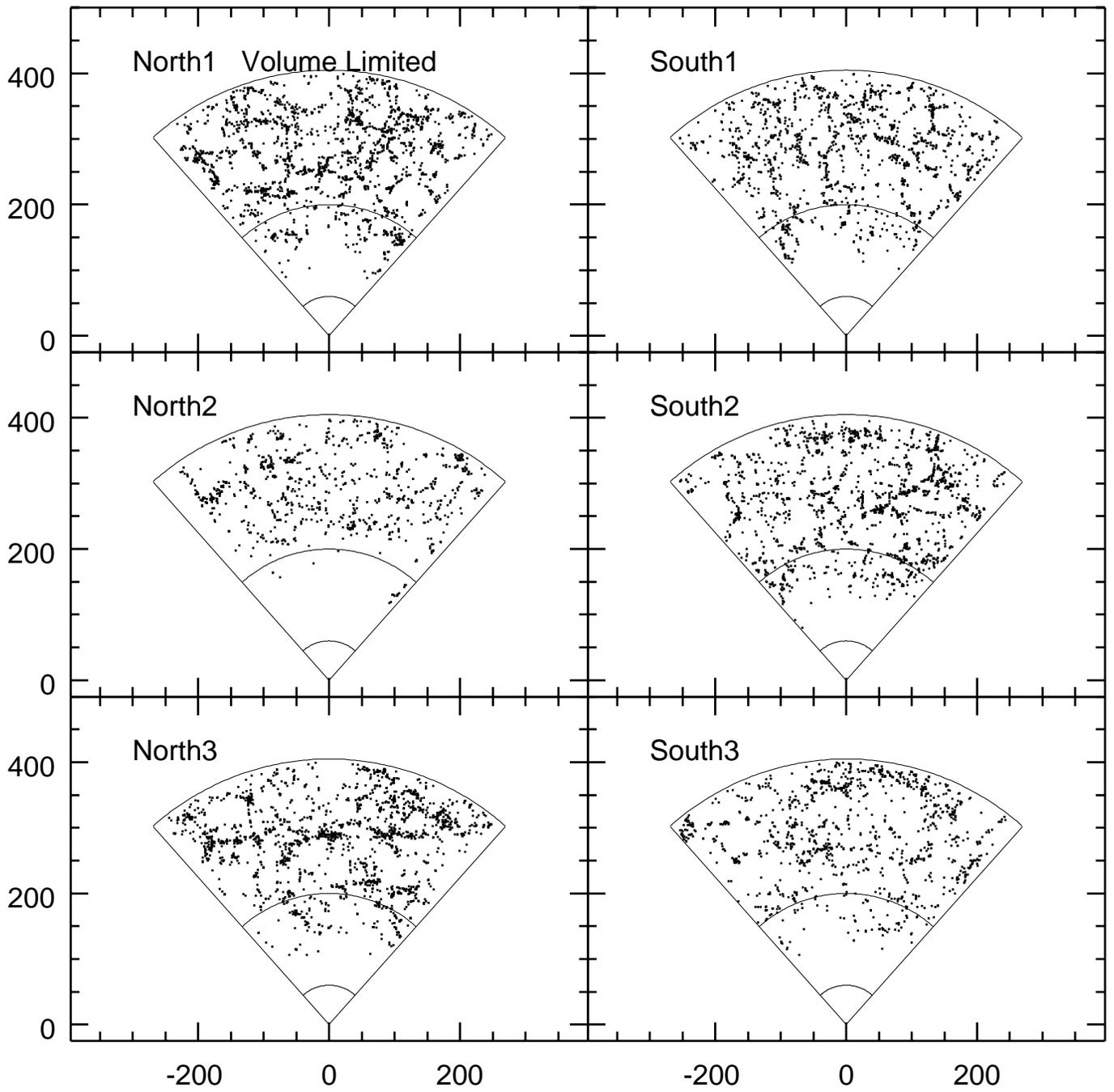
\end{document}